# SURVEY OF CYBER VIOLENCE AGAINST WOMEN IN MALAWI

Donald Flywell Malanga, University of Livingstonia, dmalanga@unilia.ac.mw

**Abstract:** The purpose of this study was to investigate the prevalence of cyber violence against women in Karonga district of Malawi. The study adopted a descriptive survey design. About 67 women participated in the survey questionnaire. The study utilised Technology Facilitated Gender-Based Violence Framework as theoretical lens. The study noted that women experienced various forms of cyber violence such as cyber bullying, cyber harassment, online defamation, cyberstalking, sexual exploitation, online hate speech, and revenge pornography. Perpetrators used digital platforms such Facebook, WhatsApp, online personal accounts, dating sites, and smartphones to carry out their evil acts. The study also found that perpetrators' motivations were driven by revenge, anger, jealousy, sexual desire and political agenda, with the intentions to harm the victims socially, psychologically, economically, and physically. The women used coping measures such as confrontation and blocking the perpetrator or by leaving the online platform. It was found that women never bothered to report the incidences to the police or community to seek for support due to lack of awareness, cultural and patriarchal factors. In conclusion, the study found that the prevalence of cyber violence against women is rising significantly in Malawi. Therefore, the findings provide insights to policy makers and research practioners on how to implement strategies for combating cyber violence against women in the country.

## 1.    INTRODUCTION

Globally millions of women are subjected to deliberate violence because of their gender (Council of Europe, 2018). Violence against women knows no boundaries, cut across boundaries, culture, race, and income groups hamming victims, people around them and the society as whole (Maundu, 2020).Violence against women is defined as any act of gender-based violence that results in physical, sexual, or mental harm or suffering to women, including threats such as coercion or arbitrary deprivation of liberty whether occurring in public or private life (World Health Organisation [WHO], 2020). Studies suggest that 1 in 3 women worldwide have experienced some form(s) of gender based violence in their life time. The growing reach of Information and Communication Technologies (ICTs) such as internet, social media, mobile devices, and digital platforms have presented potential opportunities and enabled efforts to address violence against women (Council of Europe, 2018; European Institute of Gender Equality, 2017; Frisen, Berne &Lunde, 2014). However, the same ICTs that are supposed to empower women, have also become tools to inflict them (Cybersafe, 2020). This form of violence is called cyber violence against women (CVAW).

### 1.1.    Definition of cyber violence against women

Since cyber violence against women is an emerging phenomenon, literature indicates a lack of consistent and standard definitions or methodologies used to conceptualise and measure it (Cybersafe, 2020). To this end, the Council of Europe (2018) concedes that there is not yet a stable lexicon or typology of offence considered to be CVAW. Many of the examples or types of CVAW are interconnected or overlapping or consists of a combination of acts. However, this study adopts a more encompassing definitions which posits that CVAW refers to a type of gender-based violence perpetrated through electronic communication and the internet (Cybersafe, 2020). This may result in physical, sexual, psychological or economic harm or suffering and may include the exploitation of the individual's circumstances, characteristics or vulnerabilities (Council of Europe, 2018). United Nations (UN) Broadband Commission (2015) indicates that CVAW is just as damaging to women as physical gender-based violence does. The report further estimates that 73% of women have endured CVAW, and are 27 times more likely than men to be harassed online.

## 1.2 Forms of cyber violence against women

Although cyber violence can affect both men and women, women experience different and more traumatic forms of cyber violence (Frisen et al., 2014). Literature provides various forms of CVAW such as cyber stalking, cyber harassment, cyber bullying, online sexual harassment, nonconsensual pornography, etc (West, 2014). United Nations Broad Commission (2015) indicates that women aged between18-24 years experienced some forms of cyber violence in their life time. They disproportionately suffer from cyber bullying, cyber harassment, cyber stalking and online sexual harassment, etc. The report further indicates that in the European Union (EU) of 28 membership, 18% of women have experienced cyber violence in form of online hate, online harassment, cyber stalking, and others since the age of 15. This prevalence rate represents 28million women. Sargent et al. (2016) found that over 50% of girls older than 13 years of age in Slovenia experienced some form of cyber violence. While there are various forms of CVAW, only the relevant ones applicable to this study were adopted as presented in Table 1.

| Forms of CVAW | Definitions/description |
| --- | --- |
| Cyber stalking | Involves repeated incidents, which may or may not individually be innocuous acts, but combined undermine the victim's sense of safety and cause distress, fear or alarm. |
| Cyber bullying | Entails receiving unwanted, offensive, sexually explicit emails or SMS messages; inappropriate, offensive advances on social networking websites or in Internet chat rooms |
| Cyber harassment | Is a harassment by means of email, text (or online) messages or the Internet |
| Online hate speech | Is a type of speech that takes place online with the purpose of attacking a person or a group based on their race, religion, ethnic origin, sexual orientation, disability, or gender |
| Online defamation | Involves online publication of a false statement about a person that results in some kind of harm, including financial losses or damage to the subject's reputation |
| Online exploitation | A person receiving sexual threats, being coerced to participate in sexual behaviour online, or blackmailed with sexual content. |
| Non-consensual pornography (revenge porno) | Involves online distribution of sexually graphic photographs or videos without the consent of the individual in the images |

**Table 1: Forms of cyber violence against women** (Cybersafe, 2020; European Institute of Gender Equality, 2017).

## 1.3 Impact of cyber violence against women

Though the available literature on the impact of CVAW is limited, it demonstrates that CVAW has social, physical, psychological, and economic impact on women if left unchecked. Gleenson and Garcia, (2014) investigated the relationship between cyber victimisation and body esteem among Swedish adolescents. The study found that 84% of adolescents reported behavioral impact, 80% a social impact, 56% a cognitive impact, and 12% experienced a physical impact. Study by West report (2014) found that 65% of women from the survey experienced some sort of psychological impact inform of anxiety and damaged self-image. Council of Europe (2018) reported that cyber violence against women undermines women's core fundamental rights such as dignity, access to justice and gender equality. The study by West report (2014) concluded that the impact of CVAW are social, physical, economic and psychological. Despite recent increased efforts and investments to curb CVAW prevalence globally, it continues to be a complex and pervasive public health and human rights concern that is being perpetuated in various forms and across multiple contexts and platforms

By the end of 2020, Africa recorded an 8% increase in ICT penetration (GSMA, 2020) compared from 20% in 2015 (UN Broadband Commission, 2015). Despite this exciting ICT penetration rates, the continent has not been immune to increased CVAW, with countries like Kenyan, Nigerian and South Africa becoming fast growing hubs of CVAW activity. In particular, recent studies reported that 33% prevalence of CVAW in Kenya and South Africa. The study also indicated that mobile phones and social media are currently the major ICT tools used to perpetuate CVAW in most emerging African countries (European Institute of Gender Equality, 2017). Though, efforts to tackle CVAW are emerging at a larger scale, it remains an extensive and widely under-reported online human rights violation African countries (Maundu, 2020).

Much of the studies focus on CVAW in developed world such as Europe and United States (Cybersafe, 2020; Council of Europe, 2018; European Institute of Gender Equality, 2017), and similar empirical studies are lacking in least developed countries like Malawi. Other studies contend that CVAW is an emerging concept in developing countries (Violence against women, 2014). As a result, most available studies are pilot studies in nature, sometimes conducted without theoretical foundations (International Center for Research on Women [ICRW], 2018). In this regard, designing preventive measures using the evidence from developed countries alone has substantial drawbacks due to geopolitical differences and other contextual factors that exist within countries including Malawi (Maundu, 2020).

In Malawi, there is paucity of studies on cyber violence against women. Much of the studies have focused much on gender-based violence on the physical space (Spotlight Initiative, 2020; Reliefweb, 2019; Settergren & Sapuwa, 2015), and have paid little or no attention in the digital space. Due to uniqueness nature of this study, findings from such existing studies cannot be generalised to prevalence of cyber violence against women in the country. The dearth in literature on CVAW, has led to lack of informed policy direction on understanding CVAW, its nature, prevalence and responsive strategies to address it. To address the lacuna of knowledge, this study attempted to pose this main research question: What is the prevalence of cyber violence against women in Malawi? To answer the main research question, the following four sub-questions were posed?

- What form(s) of cyber violent behaviours that women experience?
- What are the perpetrators' motivation to commit cyber violence against women
- What are the impacts of cyber violence against women?
- What strategies are available to tackle cyber violence against women?

The study drew on Technology-Facilitated Gender Based Violence Framework (TFGBVF) as theoretical lens. Using TFGBVF helped to understand the prevalence of CVAW in Malawi, with a view to increasing its awareness, nature, impacts, and advocate for personal, community and national responses to prevent and mitigate this nature of online aggression. In addition, employing TFGBVF as a theoretical framework for this study led to better conceptualisation of cyber violence study phenomena, thereby contributing to theory and practice in Information and Communication Technology for Social Justice (ICT4SJ) space. The rest of the paper is arranged as follows: background to the study, theoretical framework, methodology, analysis of findings, and discussion and conclusion.

## 2.      BACKGROUND TO THE STUDY

Malawi gained its independence from Great Britain in 1964. It borders Tanzania, Zambia and Mozambique. The country has an estimated population of 17.7 million people of which 85% live in rural areas. It is regarded as one of the least developed countries in the world with a Gross Domestic Capital per Capita is USD 516.80(WHO, 2020). Most women are working in the agricultural sector, which is a backbone of Malawi's economy (Malanga & Chigona, 2018). Of those in non-agricultural waged employment, 21% are women and 79% are men, and the numbers have remained the same over the years (Malanga, 2020). The overall mobile penetration is estimated at 45.5% while internet is 6.5% (Telecommunication Union, 2020; MACRA, 2015). About 34.5% of women own a mobile phone, 0.6% own a desktop computer, 1.8% own a Laptop, while just 4.7 % of them have access to the internet. The low rate of ICT penetration in Malawi is attributed to the country's weak economy, high value added tax (VAT) imposed on importation of ICT gadgets, and other contextual factors (Malanga& Jorosi; Malanga, 2017).

## 2.1 Gender-based violence situation in Malawi

Section 24 of the Malawi's Constitution stipulates that "women and girls have the right to full and equal protection by the law, have the right not to be discriminated against on the basis of their gender or marital status". These rights are also operationalized in Malawi's Gender Policy (2015) and National Action Plan to Combat Gender-based Violence in Malawi (2014-2020) (Settergren & Sapuwa, 2015). Recently, the Malawi enacted Electronic Transaction and Cybersecurity Act of 2016 and Data Protection Bill (2021) has also been drafted. These legislative frameworks are necessary to ensure that women's digital rights are protected online. Despite these policy interventions, gender-based violence remains high in Malawi (Malanga, 2019). In addition, undocumented reports that cases of CVAW in the country is rising as more women are interring in the digital space. The root causes point to culture and unequal power relations between men and women, which ensure male dominance over women (Spotlight Initiative, 2020). The unequal status of women is further exacerbated by poverty and discriminatory treatment in the family and public life. As a result, Malawi is ranked 173 out of 188 countries on the UN's Gender Inequality Index (USAID, 2019). Therefore, it was critical to undertake the current study to understand the prevalence cyber violence against women in Malawi so that the findings could offer policy suggestions to address the vice.

## 3.        THEORETICAL FRAMEWORK

This study was guided by Technology-Facilitated Gender-Based Violence (TFGBV) Framework as a theoretical lens (International Center for Research on Women [ICRW], 2018).The framework attempts to explain two stages how gender-based cyber violence against women occurs: From perpetrator and Survivor/Victim's perspectives.

From the perpetrator's view, the framework suggests the motivations and intentions of why perpetrators/attackers commit cyber violent behaviours against women (Cybersafe, 2020). Motivation refers to the emotional, functional, or ideological drivers behind the perpetrator's behaviour. Motivations can be political or ideological in nature driven by revenge (ICRW, 2018). From motivation comes intent. Intent is a determination of the perpetrator to harm the victim. Intent also varies by type of behaviour which may include psychological, physical harm or enforcement of gender norms. Next, the perpetrator conducts a cyber violent behaviours which could be inform of stalking, defamation, sexual harassment, exploitation and hate (ICRW, 2018). Furthermore, the framework explains the frequency, mode and cross-cutting tactics used by perpetrators or attackers to commit cyber violence against women victims/survivours (ICRW, 2018).

Likewise, on the survivor/victim's perspective, the framework postulates the impacts of cyber violence on the victim/survivor. The impact can be physical, psychological, functional, social and economic ones. The framework further suggests that the impact of cyber violence depends on the relationship that exists between the perpetrator and the victim/survivor (ICRW, 2018). The relationship can be personal, impersonal or institutional. Most importantly, the theory suggests the help-seeking and coping mechanism that should be put in place for preventing or dealing with victims/survivors of cyber violence (Cybersafe, 2020). The help-seeking and coping mechanisms can be individual, community/social/legal or national support services. To this end, this framework illustrates the range of experiences from the motivations of the perpetrators to the impact and help-seeking behaviours of the victims or survivors. This process is set within the larger context, because what constitutes cyber violence is locally defined and experienced (Cybersafe, 2020; ICRW, 2018; West, 2014). Based on the evidence, this framework was selected relevant to guide this study. Figure 1 provides an illustrative diagram that summaries the concepts of the TFGBV framework.

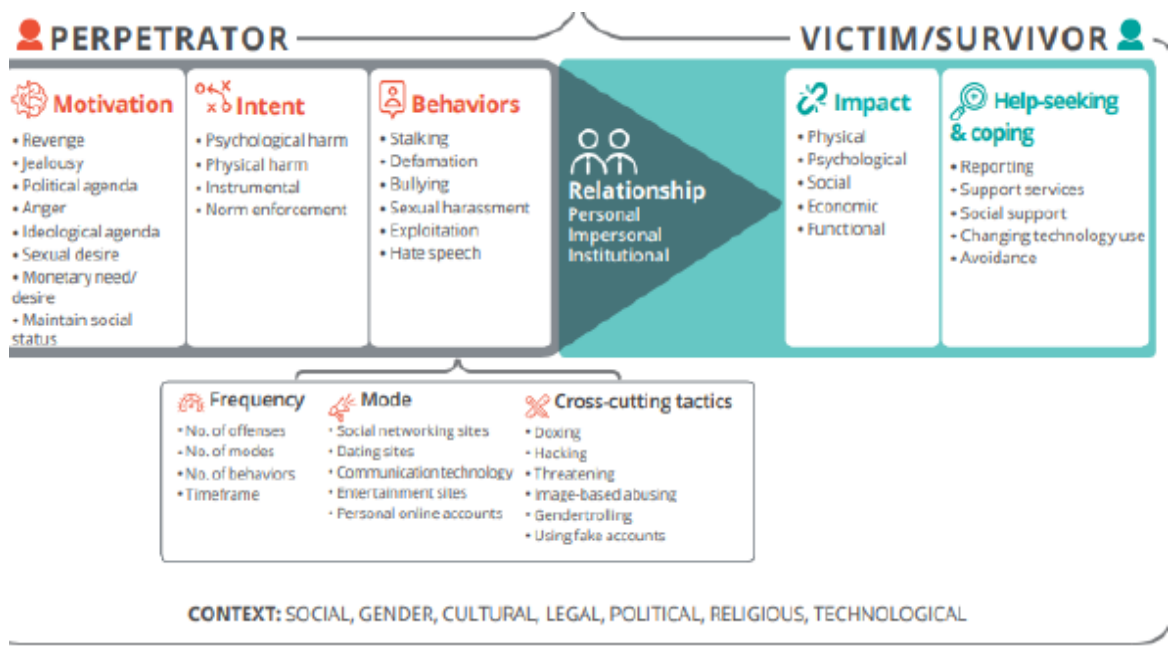

**Figure 1: Technology Facilitated Gender Based Violence Framework** (ICRW, 2018)

## 4. METHODOLOGY

### 4.1 Research design

The study adopted a cross-sectional survey design. The reasons for employing survey design were that it is popular and allows the collection of a large amount of data from sizeable population in a highly economical way (Saunders et al. 2014).

### 4.2 Population, Sampling and site selection

This study was conducted in a town market of Karonga district. This a market place where the majority of low-income merchants including women, are given spaces by the district council authorities to trade their small-scale businesses. Karonga district is situated 220km North of Mzuzu city and 50km south of Songwe border with Tanzania (Malanga & Kamanga, 2019). It has a population of 365,028 representing 1.3% of the country's population.

Women accounts for 51.7% of the population. Number of households is estimated at 74,953 in the district (Malawi Demographic and Health Survey [MDHS, 2018]. The majority of women are involved in subsistence agricultural activities as their main source of household livelihoods. Household ownership includes TVs, radios, landline, satellite dishes, mobile phones, computers and refrigerators, etc (MDHS, 2018). About 80 survey questionnaires with closed-ended questions were randomly distributed to low-income women merchants, aged between 18-45 years old. However, only 67 questionnaires were returned and suitable for analysis. The questionnaire was pre-tested to 8 women administered by the researcher himself. Pre-testing enabled the researcher to correct all ambiguities that were found in the survey questionnaire (Creswell, 2014; Saunders et al., 2014). This ensured that data collection tools were reliable and valid, before they were finally administered to the sampled population of women.

### 4.3 Research ethics, data collection and analysis

Permission to conduct the study was obtained from Karonga district council. The participants were informed that their involvement in the study was voluntary. To ensure that participants' privacy and confidentiality are protected, all recordings and transcripts were stored securely. Pseudonyms were used to protect their identities. In addition, individual written and oral consent was obtained from the respondents. Data was collected between 20ᵗʰ July to 5th August, 2020. All Covid-19 preventive measures were adhered to. The collected data was analysed descriptively using Microsoft Excel 2013 to obtained tables, frequencies and percentages for presentation of data.

## 5.    ANALYSIS OF FINDINGS

### 5.1 Demographic profile of respondents

In terms of demographic profile of respondents, the findings showed that majority of respondents (35.8% or 24) were aged between 18-25 followed by those who were aged between 26-35 (28.4% or19).  In education, 41.8% (28) attained primary school, 34.3 % (23) attained secondary school, while few (19.9% or 10) attained tertiary education. In terms of occupation, 31.3% (21) of women were students, 28.4% (19) were farmers, 17.9%(12) were employed women. Only a few of them were business owners (14.9% or 10). For marital status, 50.7% (34) were single women, 38.8% (26) women were married, and only 6.0% (4) were widows. From this analyses, it evident that the substantial number of sampled women were aged betwee18-26 and had attained primary school.

### 5.2. Forms of cyber violence women experience

Sampled women were asked about the forms of cyber violence they experienced.  As presented in Table 2, results showed that cyber stalking (92.5%), cyber bullying (83.6%), cyber harassment (76.1%), and online sexual exploitation (71.4%) are major forms of gender-based cyber violent behaviours women experience. Few respondents also experienced online hate speech, non-consensual pornographic materials, and online defamation.

| Women's  experience(s) of gender-based cyber violence | Frequency(multiple response) | Percentage |
|---|---|---|
| cyber harassment | 51 | 76.1 |
| cyber bullying | 56 | 83.6 |
| cyber stalking (e.g. false accusations, threats etc) | 62 | 92.5 |
| online hate speech | 31 | 46.3 |
| online sexual exploitation | 48 | 71.6 |
| non-consensual pornography | 36 | 53.7 |
| online defamation | 29 | 43.3 |
| others | 13 | 19.4 |

**Table 2: Showing form(s) of cyber violence women experience**

Participants were also asked to state the period or frequency, in which they usually experienced cyber violence. The findings showed that 67.1% of respondents stated that they experienced daily one or more form of gender-based cyber violence, 26.4% indicated that that gender-based cyber violence occurred weekly, while 4.5% of respondents indicated that that did not know. Respondents were also asked to state the ICT mode that attackers or perpetrators used to commit cyber violent behaviours. The majority of participants indicated that social media (62.5% or 42) such as Facebook and WhatsApp, personal online accounts (31.3% or 21) were the most frequently digital platforms used by perpetrators to commit these vices. On the other hand, entertainment and dating sites came the least. It was also revealed that perpetrators used hacking (19.7%), gender trolling (54.2%), fake accounts (33.1%), and doxing (12.3%), and communication threats (4.2%) as main tactics to gain entry into women and girls online spaces.

### 5.3. Perpetrators' motivation for committing cyber violence

Another objective was to establish the motivations behind the perpetrators to carry out cyber violence against women. As shown in Table 3, participants indicated that perpetrators' motives were inspired by revenge (83.1%), anger (76.4%), jealousy (69.3%), sexual desire (59.7%), and political agenda (32.8%). The intentions were to psychologically (61.8%) and physically (38.5%) harm the victims or survivors.

| Motives | Frequency(multiple responses) | Percentage |
|---|---|---|
| Jealousy | 46 | 69.3 |
| Sexual desire | 40 | 59.7 |
| Revenge | 56 | 83.1 |
| Political agenda | 22 | 32.8 |
| Anger | 51 | 76.4 |
| Monetary desire/agenda | 14 | 21.5 |
| Maintain social status | 12 | 17.9 |
| Ideological agenda | 8 | 12.1 |

**Table 3: Perpetrators' motivation for committing CVAW**

## 5.4. Impact of cyber violence against women

Another important objective was to find out from the respondents on how cyber violence experience impacted them socially, psychologically, economically, and physically. As Table 4 indicates, findings revealed that sampled participants withdrew from online activity (68.7%), lost reputation (17.2%), and cut down from social activity (9.0%).

| Impact of gender-based cyber violence on women | Frequency (N-67) | Percentage |
|---|---|---|
| Social impact | | |
| Harm reputation | 12 | 17.9 |
| Withdrew from online activity | 46 | 68.7 |
| Isolated from family, friends or co-workers | 2 | 3.0 |
| Cut down from on social activity | 6 | 9.0 |
| Moved out of the community | 1 | 1.5 |
| Psychological/Emotional impact | | |
| Anxiety | 6 | 9.0 |
| Living in a state of fear | 33 | 49.3 |
| Depression | 2 | 3.0 |
| Self-image damaged | 21 | 31.3 |
| Self-harming behaviours | 1 | 1.5 |
| Thoughts of suicide | 1 | 1.5 |
| Negative impact on job/school performance | 3 | 4.5 |
| Economic/Financial impact | | |
| Loss of income | 51 | 76.1 |
| Loss of educational opportunities | 4 | 6.0 |
| Loss of home | 5 | 7.5 |
| Inability to get a new job | 6 | 12.0 |
| Loss of property | 1 | 1.5 |
| Physical impact | | |
| Self-harm | 3 | 4.5 |
| Physical abuse exacerbated by online violence | 36 | 53.7 |
| Physical harm and injury resulting from online violence | 23 | 34.3 |
| Physical illness | 5 | 7.5 |

**Table 4: Showing impacts of cyber violence against women**

Participants were also abused online (53.7%) and harmed online (34.3%), which to some extent led to physical illness (7.5%). Furthermore, the findings found that cyber violence negatively impacted women and girls such as living in a state of fear (49.3%), self-image damaged (31.3%), and anxiety (9.0%). Other women stated that gender-based cyber violence has led them to loss of income (76.1%) and inability to get new employment opportunities (12.0%). Overall, this implies that cyber violence had a negative social, physical, psychological, and economical impact on women.

**5.5 Responses/strategies to tackle cyber violence against women**

Another objective of the study was to establish the responsive strategies women used to tackle cyber violence. The results are presented in Table 2. In terms of seeking social support services, the study found that half of the respondents blocked perpetrators on digital platforms(50.7%), while others left the digital platform (26.9%) or confronted the perpetrator physically or digitally (10.4%).

| Responses/strategies | Frequency (N=67) | Percentage |
|---|---|---|
| Seeking social support services | | |
| Confronted attacker/perpetrator (s) | 7 | 10.4 |
| Blocked attacker/perpetrator(s) on digital platforms | 34 | 50.7 |
| Publicized attacker(s) personal information online | 1 | 1.5 |
| Left the digital/online platform(s) | 18 | 26.9 |
| Exposed the attacker/perpetrator(s) to their family, friends, and employers | 2 | 3.0 |
| Shared information with both print and online media (newspapers, blogs, TVs, radios etc) | - | - |
| Left to a transition place/house/refuge | 1 | 1.5 |
| Sought for health/social counselling services | 4 | 6.0 |
| Seeking legal support services | | |
| Reported to the police (attacker/perpetrator (s) arrested) | 2 | 3.0 |
| Reported to the police(police took no action) | 21 | 31.3 |
| Filed civil law suit against the perpetrator/attacker | 1 | 1.5 |
| Reported to the police (attacker/perpetrator convicted) | 1 | 1.5 |
| Never reported to police/community leaders | 42 | 62.7 |
| Seeking Intervention from digital/online platform (Facebook, YouTube, pornography site, etc.) | | |
| Digital/online platform blocked the attacker/perpetrator from using the platform | 2 | 3.0 |
| Digital/online platform removed the content | 5 | 7.5 |
| Appealed to digital/online platform but platform took no action | 9 | 13.3 |
| Never sought intervention from the digital/online platform owners | 51 | 76.1 |

**Table 4: Responses/strategies for tackling cyber violence against women**

On seeking legal services, the findings showed that 62.7% of the respondents never reported to the police or community leaders, while 31.3% of respondents reported the incidences to police, but unfortunately the police took no action. However, when it comes to seeking intervention from online digital platforms, the results show that 76.1% of surveyed participants never sought any intervention from the online platform companies, while 13.3% appealed to online platforms, but the online/digital platform took no action.

# 6.    DISCUSSION AND CONCLUSION

This study aimed at investigating the prevalence of cyber violence against women in Malawi. The study targeted low-income women in of the market township of Karonga district in Malawi. Specifically, the study looked at the forms of cyber violence women experienced, the motives orchestrated by the perpetrators to commit such cyber violent behaviours, the impact that women suffered against cyber violence, and possible responsive strategies to tackle cyber violence against women.

West (2014) posits that women experience various form of cyber violent behaviours such as stalking, harassment, and bullying, sexual exploitation, among others. These violent behaviours are done by perpetrators and may be repeated with varying frequency. Perpetrators usually conduct cyber violent behaviours against women and girls using ICT and digital media such as mobile phones, social networking sites, email accounts, internet, etc (ICRW, 2018; Dredge et al. 2014).

Likewise, the present study found that women experienced cyber violent behaviours that ranged from cyber bullying, online defamation, cyber stalking, sexual exploitation to hate speech, and non-consensual pornography. Perpetrators used digital platforms such as Facebook, WhatsApp, online personal accounts, dating sites, and smartphones to carry out their acts of cyber violent behaviours against women. This was not surprising considering that the number of women accessing and using such digital platforms in Malawi has increased significantly over the past five years (Telecommunication Union, 2020). Prior studies indicate that gender-based cyber violence is informed by the connection or relationship that exists between the victim/survivor and the attacker/perpetrator (ICRW, 2018; West, 2014). In this study, it was also found that most of the perpetrators were known to the women victims inform of ex-husband, boyfriend and co-workers.

ICRW (2018) defines perpetrator's motivation as the emotional, psychological, functional or ideological drivers behind the perpetrators behaviour. In this regard, motivation can be political, ideological in nature or driven by revenge. Similarly, this study revealed that perpetrators' motivation to commit acts of cyber violence against women, were aimed at revenging, anger, jealousy, sexual desire and political agenda, with the intentions to psychologically and physically harm the victims. Similar studies have reported that perpetrators commit cyber violence against women with the intentions of harming them physically or psychologically. Other intentions are to enforce certain gender norms or extortion (Cybersafe, 2020; ICRW, 2018; Sargent et al. 2016)).

 Literature suggests that every victim or survivor of cyber violence is impacted in one way or the other by their experience.  Those impacts can bring harm to their physical, mental, social status and economic opportunities, or sometimes may lead to death (ICRW, 2018).  Likewise, the current study found that women victims or survivors were negatively impacted by cyber violence. These included social, psychological, economical, and physical impacts. In terms of social impacts, the study found that the majority of sampled women withdrew from online activity, lost reputation, and opted to cut down from social activity.  Psychologically, the study found that cyber violence experience made women start living in a state of fear, with anxiety and self-image damaged. Women were also impacted physically inform of online harm which led to physical illness to some extent.

Furthermore, the study found that some women suffered economically as a result of cyber violence. For instance, a substance number of sampled women lost their income and were unable to get new employment or participate on new business opportunities. The findings further confirm similar studies that have reported that if cyber violence against women remains unchecked could have a catastrophic impact on victims or survivours' social, economic and psychological well-being (Cybersafe, 2020; Council of Europe, 2019; West, 2014). Consequently, this is tantamount to the African Declaration on Internet rights and Freedoms (AfDec) which advocates for rights and social justice for women online (AfDec, 2015). AfDec posits that online spaces should provide safer environment for women and other marginalised groups to participate meaningfully in order to realise their social, political, economic and empowerment outcomes.

Prior studies indicate that victims/survivors of cyber violence can report their experiences to the police, seek health/psychological counselling or legal support services, and get help from their social networks (ICRW, 2018). In terms of seeking social support, the study found that women blocked perpetrators on digital platforms, left the digital platform or confronted the perpetrator physically or digitally. On seeking legal services, it was revealed that sampled women never reported cyber violent incidences to the police or community leaders. Only few of them did report the issues at the police, unfortunately, the police took no action. Moreover, seeking intervention from online digital platforms, the study found that the majority of sampled women never sought any intervention from the online platform companies, and only a few appealed to online platforms, but the online/digital platform took no action.  The findings therefore was an indication that women avoided their participation of digital spaces for fear of experiences cyber violence. It is also evident that women never bothered to seek social, community and legal support services due to lack of awareness of such support services (UN Broadband Commission, 2015). Due to systemic cultural and other patriarchal factors, women in Malawi like their counterparts in African countries, are often reluctant to report their online victimizations for fear of social repercussions (Spotlight Initiative, 2020; European Institute of Gender Equality, 2017).

In conclusion, this study has demonstrated the prevalence of cyber violence, forms of cyber violence women experience, and its nature of manifestations. Furthermore, the study has shown the impacts of cyber violence on women victims and survivors and ways of tackling it. The study has revealed forms of cyber violence against women including bullying, defamation, and stalking, sexual exploitation; hate speech, and non-consensual pornographic videos and images, etc. The findings have also shown that cyber violence against women impacted them psychologically, socially, economically and physically. Women victims or survivors of cyber violence never bothered to seek social, community or legal support as coping mechanism due to lack of awareness, cultural, and other patriarchal factors. As a result, this made women refrain from participation on online space. This is against the principles of promoting digital rights and social justice. Therefore, the study findings have both theoretical and practical implications.

- This study contributes to the Information and Communication Technology for Social Justice (ICT4SJ) space. The use of Technology Facilitated Gender Based Violence framework as a theoretical lens, provided a better theoretical and analytical descriptions of the study phenomena.

- The government should formulate strategies for addressing cyber violence against women. The voices of women who are victims of the phenomena

- Government, civil society and other private actors should advocate for awareness-raising campaigns for women about cyber violence. This awareness and training might help women get empowered and make decisions on where to seek social or legal support, in case they have been abused online.

- Online human rights advocates can use evidence generated from this study to inform and gain supportive campaigns and call for legal protections of women abused or harmed against cyber violence.

- Internet and other online platforms such as Facebook should create clear options for getting online images or abusive content removed (Council of Europe, 2018). They should also respond immediately and effectively to complaints from victims of online abuse, and finally establish genuine consent for terms of use.

Despite the empirical contribution of this study makes to ICT4JS research, it has some limitations. This study focused only on one market township of Karonga district. Therefore, future studies should be replicated in other districts so that the findings could be generalised. The study also utilised a small sample of women, utilising descriptive survey design which has its own weakness. Future studies could utilise a bigger sample with mixed method, covering women from both rural and urban areas. This could help to gain more insights into the study phenomena. The study has also shown that prevalence of cyber violence against women is increasing significantly in the country. This study therefore, provides evidence that policy makers, research practioners and other stakeholders can utilise to put in place urgent responsive strategies to combat cyber violence in the country.

# REFERENCES AND CITATIONS